\documentstyle[proceedings,psfig,numreferences]{crckapb}

\begin{opening}
\title{LOW-ENERGY DIFFRACTION; A DIRECT-CHANNEL POINT OF VIEW: THE BACKGROUND}

\author{\underline{L.L.~JENKOVSZKY}}
\institute{N.N. Bogolyubov Institute for Theoretical Physics\\ 
Academy of Sciences of Ukraine\\
Metrolohichna 14b, 03143 Kiev, Ukraine}
\author{S.Yu.~KONONENKO}
\institute{Physics Department, T.H. Shevchenko National State University\\
Academician Glushkov avenue 6, 03022 Kiev, Ukraine}
\author{V.K.~MAGAS}
\institute{N.N. Bogolyubov Institute for Theoretical Physics\\ 
Academy of Sciences of Ukraine\\
Metrolohichna 14b, 03143 Kiev, Ukraine $\&$ \\
Center for Physics of Fundamental
Interactions (CFIF)\\
Physics Department, Instituto Superior Tecnico\\
Av. Rovisco Pais, 1049-001 Lisbon, Portugal}

\end{opening}

\runningtitle{Low-energy diffraction: the background}

\newcommand{\beq}{\begin{equation}}
\newcommand{\eeq}[1]{\label{#1} \end{equation}}
\newcommand{\insertplot}[1]{\centerline{\psfig{figure={#1},width=13.0cm}}}

\begin{document}


\begin{abstract}
We argue that at low-energies, typical of the resonance region,
the contribution from direct-channel exotic trajectories replaces
the Pomeron exchange, typical of high energies. A dual model
realizing this idea is suggested. While at high energies it
matches the Regge pole behavior, dominated by a Pomeron exchange,
at low energies it produces a smooth, structureless behavior of
the total cross section determined by a direct-channel nonlinear exotic
trajectory, dual to the Pomeron exchange.
\end{abstract}

In this paper we investigate the role of the low-energy background
in diffractive processes. We follow the ideas of two-component
duality \cite{F}, \cite{H}, by which the high-energy Pomeron
exchange is dual to the low-energy background. In describing this
background we use a dual amplitude with Mandelstam analyticity
(DAMA) \cite{DAMA}, where, contrary to narrow resonance dual
models, nonlinear trajectories are not only allowed, but even
required by their general properties. The asymptotic rise of the
trajectories in DAMA is limited by the condition $
|{\alpha(s)\over{\sqrt s\ln s}}|\leq const, \ \
s\rightarrow\infty.$ 
The most popular trajectories, satisfying above condition,
are square root type of trajectories \cite{DAMA} , used also in this work.  
Such trajectories have an upper bound on the real part of
these trajectories, which results in the termination of the
resonances on it. The maximal value of this real part is
determined by the free parameters of the trajectories, that can be
fitted to the resonance spectra.

An extreme case is that of an
exotic trajectory, whose real part does not reach any resonance.
Relevant examples are presented in this talk.

The $(s,t)$-term of DAMA is
\beq 
D(s,t)=\int_0^1 {dz \biggl({z \over g}
\biggr)^{-\alpha(s')-1} \biggl({1-z \over
g}\biggr)^{-\alpha(t')-1}}\ , 
\eeq{eq1}
where $s'=s(1-z)$, $t'=tz$,
$g$ is a parameter, $g>1$, and $s$, $t$ are the Mandelstam
variables.

For $s\rightarrow\infty$ and $t=0$ it has the following Regge
asymptotic behavior 
\beq
D(s,t)\approx\sqrt{{2\pi\over{\alpha_t(0)}}}g^{1+a+ib}\Biggl({s\alpha'(0)g\ln
g\over{\alpha_t(0)}}\Biggl)^{\alpha_t(0)-1}\ , 
\eeq{eq2} where 
$a=Re\ \alpha\Bigl({\alpha_t(0)\over{\alpha'(0)\ln g}}\Bigr)$ and 
$b=Im\ \alpha\Bigl({\alpha_t(0) \over{\alpha'(0)\ln g}}\Bigr)$.

The pole structure of DAMA is similar to that of the Veneziano
model except that multiple poles may appear at daughter levels.
The presence of these multipoles does not contradict the
theoretical postulates.  On the other hand, they can be removed
without any harm to the dual model by means the so-called Van der
Corput neutraliser. The procedure  \cite{DAMA} is to multiply the
integrand of (\ref{eq1}) by a function $\phi(x)$ with the properties: $$
\phi(0)=0,\ \ \ \phi(1)=1,\ \ \ \phi^n(=0),\ \ n=1,2,3,... $$ The
function $$ \phi(x)=1-exp\Biggl({-x\over{1-x}}\biggr), $$ for
example, satisfies the above conditions and results \cite{DAMA} in a
standard, "Veneziano-like" pole structure:
\beq
D(s,t)=\sum_ng^{n+\alpha_t(0)}{C_n\over{n-\alpha(s)}}\ ,
\eeq{eq3}
where $C_n$ are the residues, whose form is fixed by the dual
amplitude \cite{DAMA}:
$$
C_n={\alpha_t(0)\Bigl(\alpha_t(0)+1\Bigr)...\Bigl(\alpha_t(0)+n+1\Bigr)\over{n!}}.
$$

The pole term in DAMA is a generalization of the Breit-Wigner formula,
comprising a whole sequence of resonances lying on a complex
trajectory $\alpha(s)$. Such a "reggeized" Breit-Wigner formula
has little practical use in the case of linear trajectories,
resulting in an infinite sequence of poles, but it becomes a
powerful tool if complex trajectories with a limited real part and
hence a restricted number of resonances are used.
If $Re\ \alpha(s)>0,$ equation (\ref{eq3}) produces a sequence of
Breit-Wigner resonances lying on the trajectory $\alpha(s)$.

Near the threshold, $s\rightarrow s_0$ 
$$
D(s,t)\simeq \frac{g^2}{\alpha(s_0)}\left(\frac{1-{s_0\over s}}{g}\right)^{-\alpha(s_0)}
Im\ \alpha(s) \cdot
$$
\beq 
\cdot \left[\ln \left(\frac{x_1\left(1-{s_0\over s}\right)}{g}\right) 
\frac{Im\ \alpha(s_1)}{Im\ \alpha(s)}+\left({1\over \alpha(s_0)-\ln x_1}\right)\right]\ , 
\eeq{a2}
where $0<x_1<1$ and $s_1=s_0+(s-s_0)(1-x_1)$.

A simple model of trajectories satisfying the threshold and
asymptotic constraints is a sum of square roots \cite{DAMA} 
\beq
\alpha(s)=\alpha_0+\sum_i\gamma_i\sqrt{s_i-s}.
\eeq{eq4}
 The number of thresholds included depends on the model; while the
 lightest threshold gives the main contribution to the imaginary part,
 the heaviest one promotes the rise of the real part (terminating
 at the heaviest threshold).

 A particular case of the model eq. (\ref{eq4}) is that with a single
 threshold. Imposing an upper bound  on the real part of this
 trajectory, $Re\ \alpha(s)<0,$ and inserting it to eqs. (\ref{eq1}) or (\ref{eq3}),
 we get an
amplitude that does not produce resonances, since the real part of
the trajectory does not reach $n=0$ where the fist pole could
appear. Its imaginary part instead rises indefinitely,
contributing to the total cross section with a sooth background.

By using the dual model (\ref{eq1}), we now calculate numerically the
total cross sections 
\beq 
\sigma_t(s)=Im\ A(s,t=0)\ , 
\eeq{eq5} where 
\beq
A(s,t)=c\left(D(s,t)+D(u,t)\right) 
\eeq{eq6} 
and $c$ is a normalization coefficient.

In this work we propose the following simplified model.\footnote{Our 
earlier attempts to model the background 
(as well as resonance contributions) can be found in Refs. \cite{JM}.}
In the $t-$ channel we use a Pomeron trajectory of the form 
\beq
\alpha_P(t)=1.1+0.2t+0.02(\sqrt{4m_{\pi}^2}-\sqrt{4m_{\pi}^2-t}),
\eeq{eq7} 
while the exotic $s-$ channel trajectory is 
\beq
\alpha_E(s)=\alpha_0+\alpha_1(\sqrt{s_0}-\sqrt{s_0-s}). 
\eeq{eq8}

In the Pomeron trajectory (\ref{eq7}) the linear term mimics high-mass
thresholds, important for the nearly exponential shape of the
cone, while the lowest, $2m_{\pi}$ threshold is manifest as a
"break" in the diffraction cone near $0.1\ GeV^2$.

By definition, the real part of the exotic trajectory should not
pass through resonances; we tentatively set
\beq
Max\{Re\ \alpha_E(s)\}=\alpha_E(s_0)=\alpha_0+\alpha_1\sqrt{s_0}=-1\ ,
\eeq{a3} 
so that it 
does not reach $n=0$, where
summation in eq. (\ref{eq3}) starts. For the lowest threshold in the exotic
trajectory we choose $s_0=(m_p+m_{J/\psi})^2,$ with $J/\psi$
photoproduction (or  $J/\psi p$ scattering) in mind.

With this exotic trajectory the threshold behaviour of   
our amplitude is:
$$
D(s,t) \sim \left(s-s_0\right)^{1/2-\alpha_E(s_0)}[const+\ln(1-s_0/s)]=
$$
\beq 
\left(s-s_0\right)^{3/2}[const+\ln(1-s_0/s)]\ . 
\eeq{lim}

The only remaining free parameters are $g, \ \  g>1$ and the slope
of the exotic trajectory $\alpha_1$. For illustrative purposes we
set $g=10$ and $\alpha_1=0.2\ GeV^{-2}$.

With these inputs in hand, now we can calculate the total cross
section for all values of $s$,  from the threshold to the highest
asymptotic values. By inserting the trajectories (\ref{eq7}) (setting
$t=0$) and (\ref{eq8}) into (\ref{eq1}), we calculate the integrals numerically.

\begin{figure}[htb]
        \insertplot{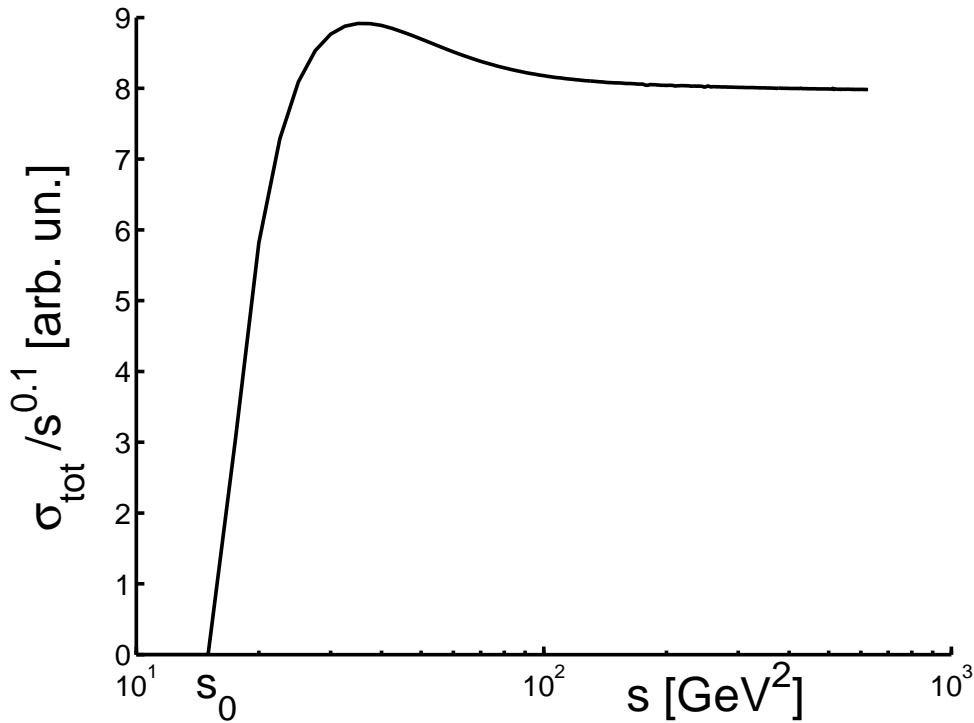}
\caption{
Purely diffractive contribution to the
total cross section (in arbitrary units) calculated from DAMA, 
eq. (\ref{eq1}), scaled to its Regge asymptotic behavior. 
The direct-channel exotic trajectory gives at low
energies non-neglectable 
contribution (background), 
which could be seen as a cusp of $\sim 10-12$ $\%$.
}
\label{f1}
\end{figure}

To see more clearly the contribution from the background, we have
divided the cross section by its Regge asymptotics $\sim s^{0.1}.$
The result, in arbitrary units, is shown in Fig. \ref{f1}. One can
clearly see a cusp between the flat asymptotic behavior and steep
rise from the threshold. This is the contribution from the
background,  modeled by the contribution from the
direct-channel exotic trajectory, amounting to about $10-12\ \%$.

The background plays an important role in the resonance region. Of
particular practical interest is the correct account for the
background in the analysis of the JLab data on the
electroproduction of nucleon resonances (see \cite{dual}).

The best testing field for this model is $J/\psi$ scattering,
where the exchange of secondary trajectories is forbidden and
consequently no resonances are expected in the low energies
region, dominated by the background, or the contribution of the
direct-channel exotic trajectory. Since $J/\psi$ total cross
section is not measured directly, the relevant reaction is
$J/\psi$ photoproduction. In most of the papers on this subject,
analyzing the HERA data on photoproduction 
(see e.g. \cite{photo} and refernces therein), the amplitude is
assumed to be Regge behaved, eventually corrected by a threshold
factor, but in any case ignoring the
contribution from the background. The evaluation of the elastic
cross section from (\ref{eq1}) however involves also integration in $t,$
thus making the calculations more complicated. This will be done
in a forthcoming paper.

{\it Acknowledgment.} We thank A.I. Bugrij for fruitful discussions on
the subject of this paper. L.L.J. and V.K.M. acknowledge the support
from INTAS, grant 00-00366.


\begin{thebibliography}{99}
\bibitem{F} Freund, P. (1968) Finite energy sum rules and bootstraps, 
{\it Phys. Rev. Lett.}, {\bf 20}, pp. 235-237.
\bibitem{H} Harari, H. (1968) Pomeranchuk trajectory and its relation to low-energy scattering amplitudes,
{\it Phys. Rev. Lett.}, {\bf 20}, pp. 1395-1398.
\bibitem{DAMA} Bugrij, A.I. et al. (1973) Dual Amplitudes with Mandelstam Analyticity, 
{\it Fortschritte der Physik}, {\bf 21}, p. 427.
\bibitem{JM} 
Fiore, R., Jenkovszky, L.L., Magas, V.  (2001) 
Connection between lepton- and hadron-induced diffraction phenomena,
{\it Nucl. Phys. Proc. Suppl.}, {\bf 99A},  pp. 131-138; 
Jenkovszky, L.L., Magas, V.K., Predazzi, E. (2001)
Resonance-reggeon and parton-hadron duality in strong interactions,
{\it Eur. Phys. J.}, {\bf A12}, pp. 361-367; 
Duality in strong interactions, nucl-th/0110085; 
Jenkovszky, L.L., TKorzhinskaja, T., Kuvshinov, V.I. and Magas, V.K. 
(2001) Duality relation between small- and large-$x$ structure functions,
Proceedings of the New Trend in High-Energy Physics, Yalta, Crimea, Ukraine,
September 22-29, 2001, edited by P.N. Bogolyubov and L.L.
Jenkovszky (Bogolyubov Institute for Theoretical Physics, Kiev,
2001), pp. 178-185.
\bibitem{dual} Fiore, R. et al. Explicit model realizing parton-hadron duality, 
hep-ph/0206027, to appear in Eur. Phys. J A.
\bibitem{photo} Fiore, R., Jenkovszky, L.L. and Paccanoni, F. 
(1999) Photoproduction of heavy vector mesons at HERA: a test for diffraction,
{\it Eur.Phys. J.}, {\bf C10}, 461-467.
\end{thebibliography}
\end{document}